\documentclass{ws-procs9x6-cpt16}

\begin{document}

\title{Nonminimal Lorentz-Violating Effects in Photon Physics}

\author{Alysson F\'abio Ferrari}

\address{Universidade Federal do ABC, Rua Santa Ad\'elia 166\\
09210-170 Santo Andr\'e, SP, Brazil}

\begin{abstract}
We study an extension of QED involving a light pseudoscalar (an axion-like
particle), together with a very massive fermion which has Lorentz-violating 
interactions with the photon and the pseudoscalar, including
a nonminimal Lorentz-violating coupling. We investigate the low energy
effective action for this model, after integration over the fermion
field, and show that interesting results are obtained, such as the
generation of a correction to the standard coupling between the axion-like
particle and the photon, as well as Lorentz-violating effects in the
interaction energy involving electromagnetic sources such as pointlike
charges, steady line currents and Dirac strings. 
\end{abstract}

\bodymatter

\phantom{}\vskip10pt\noindent
The Standard-Model Extension (SME) can be understood as an effective
field theory derived from a more fundamental theory that lives in
some high energy scale, presumably the Planck energy $E_{P}$.
Therefore, it is a natural expectation that the SME should contain
minimal (renormalizable) Lorentz-violating (LV) operators, as well
as an infinite number of nonminimal, higher derivative, nonminimal
operators. On dimensional grounds, the latter are expected to be suppressed
by inverse powers of the high energy scale $E_{P}$, so they should not
be the most relevant at low energy. Indeed, until recently, most investigations 
were restricted to the set of minimal LV operators due to these general
expectations, apart from the obvious fact that this restriction reduces
the infinite set of nonminimal LV operators to a still large but
finite set of operators. This approach proved itself very
fruitful, yielding an impressive set of constraints on LV from a wide
variety of experiments and observations.\cite{Kostelecky:2008ts}

A natural question is therefore whether nonminimal LV operators\cite{km09,km13} 
could induce observable effects in low energy physics. Finding specific
instances of low energy phenomenology where a nonminimal LV operator
could give a contribution not obscured by another minimal
LV operator would be particularly interesting. In Ref.\ \refcite{ALP},
we studied a particular model that realizes this scenario, where the
presence of a very massive fermion field, with specific LV interactions,
induces at low energy a correction to the very
same coupling between light pseudoscalars and photons that is subject
to several experimental investigations. The QCD axion\cite{Peccei:1977hh}
is a specific example of this class of light pseudoscalars, 
but our discussion is valid for general
axion-like particles (ALPs).\cite{Ringwald:2012cu} 

An ALP is represented in QFT by a pseudoscalar field $\phi$ and couples to the photon via the interaction
\begin{equation}
\mathcal{L}_{{\rm ALP}}=g_{\phi\gamma}\phi{\bf E}\cdot{\bf B}\,,\label{eq:1}
\end{equation}
which allows for searches for ALPs in
low energy photons experiments. To obtain an LV correction to Eq.\ (\ref{eq:1}),
we postulated the existence of a very massive fermion field $\psi$
that couples both to the photon and to the ALP, 
\begin{equation}
\mathcal{L}=\mathcal{L}_{A}+\mathcal{L}_{\phi}+\mathcal{L}_{\psi}-\bar{\psi}\left[\gamma^{\mu}(qA_{\mu}+F_{\mu\nu}d^{\nu})+\gamma_{5}\gamma^{\mu}b_{\mu}\phi\right]\psi\,,\label{eq:2}
\end{equation}
where $\mathcal{L}_{A}$, $\mathcal{L}_{\phi}$, and $\mathcal{L}_{\psi}$
denote the standard free lagrangians for the photon, the ALP, and
the fermion field, respectively. We notice in Eq.\ (\ref{eq:2})
the presence of a minimal LV coupling $b^{\mu}$, as well as a non
minimal one, $d^{\mu}$. Integration of the fermion field yields a
low energy effective lagrangian which contains the term
\begin{equation}
\mathcal{L}_{eff}\supset Cq\thinspace b_{\lambda}d^{\kappa}\epsilon^{\rho\mu\nu\lambda}\thinspace\phi F_{\rho\mu}F_{\kappa\nu}=2Cq\thinspace\left(b\cdot d\right)\thinspace{\bf E}\cdot{\bf B}\thinspace.\label{eq:3}
\end{equation}
The coefficient $C$
is finite, but regularization dependent. The Feynman integrals contributing
to Eq.\ (\ref{eq:3}) are actually the same studied some years ago
regarding the perturbative generation of the CFJ lagrangian.\cite{CFJ,JackivKostel,Brito:2007uc}
From the experimental viewpoint, this result means that the ALP-photon
interaction measured by experimentalist could contain a hidden LV
contribution, arising from the specific setting of high energy LV
studied by us.

Motivated by these results, we also calculated the complete correction
induced in the low energy photon sector of our model, due to the presence
of the massive fermion. Disregarding the presence of the pseudocscalar
$\phi$ in Eq.\ (\ref{eq:2}) and after integration over the fermion
field, we obtained\cite{Highoperators} 
\begin{multline}
S_{eff}\left[A\right]=-\frac{1}{48\pi^{2}}\int d^{4}x\,\Biggl\{-\ln\left(\frac{M^{2}}{\mu^{2}}\right)\widetilde{F}^{2}+\frac{1}{8M^{4}}\Biggl[\frac{2}{15}\mathrm{Tr}\left(\widetilde{F}^{4}\right)-\frac{1}{3}\widetilde{F}^{4}\\
-\frac{1}{2}\left(^{*}\widetilde{F}^{\mu\nu}\widetilde{F}_{\mu\nu}\right)^{2}\Biggr]+\frac{1}{16M^{8}}\widetilde{F}^{2}\left[\frac{2}{5}\mathrm{Tr}\left(\widetilde{F}^{4}\right)+\frac{1}{2}\left(^{*}\widetilde{F}^{\mu\nu}\widetilde{F}_{\mu\nu}\right)^{2}\right]\\
+\frac{5}{96M^{12}}\left[\widetilde{F}^{4}-\left(^{*}\widetilde{F}^{\mu\nu}\widetilde{F}_{\mu\nu}\right)^{2}\right]\left[\frac{2}{5}\mathrm{Tr}\left(\widetilde{F}^{4}\right)+\frac{1}{2}\widetilde{F}^{4}\right]+\cdots\cdots\Biggr\}\,,\label{eq:seffpronta}
\end{multline}
where $M$ is the mass of the fermion, and
\begin{equation}
\widetilde{F}_{\mu\nu}=qF_{\mu\nu}+d^{\lambda}\left(\partial_{\mu}F_{\nu\lambda}-\partial_{\nu}F_{\mu\lambda}\right)\ .\label{LV}
\end{equation}
This effective action incorporates all the effects arising in low
energy photon physics due to the presence of the massive fermion $\psi$
and the LV coupling $d^{\mu}$. It contains both nonlinear effects,
starting with the so-called Euler-Eisenberg term,
\begin{equation}
{\cal L}_{F^{4}}=\frac{R}{8\pi}\left({\bf E}^{2}-{\bf B}^{2}\right)^{2}+\frac{S}{8\pi}\left({\bf E}\cdot{\bf B}\right)^{2}\thinspace,\label{eq:LEH}
\end{equation}
where $R={q^{4}}/{45\pi M^{4}}$ and $S=7R$, as well as LV
corrections, whose leading term is
\[
{\cal L}_{LV}^{\left(1\right)}=\frac{g}{12\pi^{2}}\ln\left(\frac{M^{2}}{\mu^{2}}\right)d_{\alpha}F_{\mu\nu}\left(\partial^{\mu}F^{\nu\alpha}\right)\thinspace.
\]

A natural question is whether wave propagation in vacuum is modified
in this model. Considering only modifications up to first order in
the nonlinearities and LV, we obtain the following set of modified
Maxwell's equations in vacuum,
\begin{equation}
{\bf \nabla}\cdot{\bf D}=-C^{\prime}q\mathbf{d}\cdot\left\{ -\nabla\times\left(\nabla\times\mathbf{E}+\partial_{0}\mathbf{B}\right)\right\} \thinspace,\label{eq:ME1}
\end{equation}
and
\begin{align}
-\partial_{0}{\bf D}+({\bf \nabla}\times{\bf H}) & =-d^{0}\nabla\times\left[\nabla\times\mathbf{E}+\partial_{0}\mathbf{B}\right]\nonumber \\
 & +\mathbf{d}\cdot\nabla\left[\partial_{0}\mathbf{E}-\nabla\times\mathbf{B}\right]-\mathbf{d}\times\left[\partial_{0}^{2}\mathbf{B}-\nabla^{2}\mathbf{B}\right]\nonumber \\
 & -\nabla\left(\mathbf{d}\cdot\left[\partial_{0}\mathbf{E}-\nabla\times\mathbf{B}\right]\right)\thinspace,\label{eq:ME2}
\end{align}
together with the unmodified equations ${\bf \nabla}\cdot{\bf B}=0$ and ${\bf \nabla}\times{\bf E}=-\partial_{0}{\bf B}$, where
\begin{align}
{\bf {D}} & =\left(1+C^{\prime}g^{2}\right){\bf {E}}+2R\left({\bf {E}}^{2}-{\bf {B}}^{2}\right){\bf {E}}+S\left({\bf {E}\cdot{\bf {B}}}\right){\bf {B}}\thinspace,\\
{\bf {H}} & =\left(1+C^{\prime}g^{2}\right){\bf {B}}+2R\left({\bf {E}}^{2}-{\bf {B}}^{2}\right){\bf {B}}+S\left({\bf {E}\cdot{\bf {B}}}\right){\bf {E}\ ,}
\end{align}
and $C^{\prime}=-\frac{1}{3\pi}\ln\left({M^{2}}/{\mu^{2}}\right)$.
It is evident that the right-hand side of Eqs.\ (\ref{eq:ME1}) and (\ref{eq:ME2})
vanish for the standard Maxwell theory. In our case, these terms are
not zero; however they are of second order in $R$, $S$ and $d^{\mu}$,
so in this approximation we have the surprising result that the LV
completely decouples from wave propagation. 

Nontrivial LV effects can, however, be found in the interaction between
electromagnetic sources. We singled out from Eq.\ (\ref{eq:seffpronta})
one specific LV operator, which allows us to calculate exactly the
classical electromagnetic interaction between sources. The
specific lagrangian we considered was
\begin{equation}
{\cal L}_{LV}^{\left(2\right)}=-\frac{1}{4}F_{\mu\nu}F^{\mu\nu}+\frac{1}{2}d^{\lambda}d_{\alpha}\partial_{\mu}F_{\nu\lambda}\partial^{\nu}F^{\mu\alpha}+J^{\mu}A_{\mu}\ .\label{eq:1-1}
\end{equation}
By fixing the Coulomb gauge and finding the exact photon propagator
$D^{\mu\nu}(x,y)$, we can calculate the interaction energy between
different sources,
\begin{equation}
E=\frac{1}{2T}\int\int d^{4}x\ d^{4}yJ_{\mu}(x)D^{\mu\nu}(x,y)J_{\nu}(y)\ ,\label{zxc1}
\end{equation}
where $T$ is the time variable, and the limit $T\to\infty$ is implicit.
In Ref.~\refcite{LVfonts} we considered several
situations involving pointlike charges, steady line currents and
Dirac strings, obtaining new effects arising from the LV. As an example,
a dipole would experience a spontaneous torque depending on the angle
between the dipole and the space component of the background vector
$d^{\mu}$. These are effects that could possibly lead to experimental
signals involving the nonminimal LV coupling $d^{\mu}$.
 
\section*{Acknowledgments}

This work was supported by the grants CNPq 482874/2013-9,
FAPESP 2013/22079-8 and 2014/24672-0.

\end{document}